\documentclass[11pt]{article}

\usepackage{latexsym,color,graphicx,amsmath,amssymb,mathtools,amsthm}

\def\bd{{\partial}}
\def\dist{{\sf dist}}
\def\RNG17{{\rm 17RNG}}
\def\GNG17{{\rm 17GNG}}
\def\Vor17{{\rm Vor}_{17}}

\newcommand{\old}[1]{{{}}}

\newtheorem*{theorem}{Theorem}

\newtheorem*{corollary}{Corollary}
\newtheorem*{lemma}{Lemma}

\begin{document}

\title{Bottleneck Matching in the Plane\thanks{%
  Work by Matthew Katz was partially supported by 
  Grant 2019715/CCF-20-08551 from the US-Israel Binational Science Foundation/US National Science Foundation.
  Work by Micha Sharir was partially supported by ISF Grant 260/18.}}

\author{Matthew J. Katz\thanks{%
  Department of Computer Science, Ben Gurion University, Beer Sheva, Israel; 
  {\sf matya@cs.bgu.ac.il}}
\and
Micha Sharir\thanks{%
  School of Computer Science, Tel Aviv University, Tel Aviv Israel;
  {\sf michas@tauex.tau.ac.il}}}

\date{}

\maketitle

\begin{abstract}
We present an algorithm for computing a bottleneck matching in a set of $n=2\ell$
points in the plane, which runs in $O(n^{\omega/2}\log n)$ deterministic time, 
where $\omega\approx 2.37$ is the exponent of matrix multiplication.
\end{abstract}

Let $P$ be a set of $n=2\ell$ points in the plane, and let $G=(P,E)$ denote the complete 
Euclidean graph over $P$, which is an undirected weighted graph with $P$ as its set of vertices,
and the weight of an edge $e=(p,q) \in E$ is $\dist(p,q)$, where $\dist$ denotes the Euclidean distance.
For a perfect matching $M$ in $G$, let $\lambda(M)$ denote the length of a longest edge in $M$. 
A perfect matching $M$ is a \emph{bottleneck matching} of $P$ if $\lambda(M) \le \lambda(M')$, 
for any other perfect matching $M'$ in $G$.

In this note, we present an algorithm for computing a bottleneck matching $M^*$ of $P$,
with running time $O(n^{\omega/2}\log n) = O(n^{1.187})$, where $\omega < 2.3728596$ 
is the exponent of matrix multiplication~\cite{AVW,LeG}.
All previous algorithms for computing $M^*$ are based on the algorithm of Gabow and Tarjan~\cite{GT},
which computes a bottlenck maximum matching in a (general) graph with $n$ vertices and $m$ edges 
in time $O((n\log n)^{1/2}m)$. These algorithms apply Gabow and Tarjan's algorithm to a subgraph 
$G'$ of $G$, with only $O(n)$ edges, that is guaranteed to contain $M^*$. The running time of
these algorithms is therefore $O(T(n) + n^{3/2}\log^{1/2} n)$, where $T(n)$ is the time needed 
to compute some suitable subgraph $G'$. In the work of Chang et al.~\cite{CTL}, $T(n)=O(n^2)$, 
and in the subsequent work of Su and Chang~\cite{SuC91}, $T(n)=O(n^{5/3}\log n)$, which is the 
best previously-known bound for computing $M^*$; see below for further details concerning 
possible choices of $G'$. In contrast to the previous algorithms, we replace the algorithm 
of Gabow and Tarjan (which immediately sets a lower bound of $\Omega(n^{3/2}\log^{1/2} n)$ 
on the running time) by an algorithm of Bonnet et al.~\cite{BCM} for computing a maximum 
matching (not necessarily bottleneck) in an intersection graph of geometric objects in the 
plane, and use it as a \emph{decision} procedure in our search for $M^*$.

We first observe that $\lambda^* \coloneqq \lambda(M^*)$ is the distance between some 
pair of points in $P$. For a real number $r > 0$, let $G_r$ denote the graph that is 
obtained from $G$ by retaining only the edges of length at most $r$. That is,
$G_r=(P,E_r)$, where 
\[
E_r = \{(p,q) \in P \times P \mid p \ne q \text{ and } \dist(p,q) \le r\} .  
\]
$G_r$ is also the intersection graph of the set of disks of radius $r/2$ centered at the
points of $P$, and by suitably scaling the configuration, it is the \emph{unit-disk graph} 
over $P$ (see~\cite{CCJ}). Bonnet et al.~\cite{BCM} presented an $O(n^{\omega/2})$-time 
algorithm for computing a maximum matching in $G_r$. Their algorithm can therefore also 
be used to determine whether $G_r$ contains a perfect matching, which is the case if and 
only if the number of edges in the returned maximum matching is $\ell$. Let $r^* > 0$ be 
the smallest value for which $G_{r^*}$ contains a perfect matching. Then, by applying 
the algorithm of Bonnet et al.~to $G_{r^*}$, we obtain a bottleneck matching $M^*$ of 
$P$ (with $\lambda^* = r^*$).

Since $r^*$ is the distance between some pair of points in $P$, we can perform a binary search in 
the (implicit) set of the $n \choose 2$ distances determined by pairs of points in $P$, using the 
algorithm of Bonnet et al.~to resolve comparisons, thereby obtaining $r^*$ and $M^*$. To perform
the binary search, we use a distance selection algorithm (such as in \cite{AASS}), whose running 
time is $O^*(n^{4/3})$ (where the $O^*(\cdot)$ notation hides subpolynomial factors), 
to find the next distance to test. Since $\omega/2 < 4/3$, we obtain a bottleneck matching 
algorithm for $n=2\ell$ points in the plane with running time $O^*(n^{4/3})$, which 
is already significantly faster than the best known such algorithm. 

However, we can do better. If we are given a set $D$ of $m$ candidate distances that includes 
$r^*$, we can obtain, in total time $O(m+n^{\omega/2}\log m)$, a bottleneck matching of $P$, 
by performing a binary search in $D$. Our goal is, therefore, to find such a candidate set $D$
of size $m = O(n^{\omega/2})$ (in fact, our set will be of size $O(n)$) that can also be generated 
efficiently. A natural choice of $D$ is the set of lengths of a suitable subgraph $G'$ of $G$
that can be shown to contain a bottleneck matching of $P$.

Chang et al.~\cite{CTL} proved that the \emph{17-relative neighborhood graph} of $P$, denoted as
$\RNG17(P)$, contains a bottleneck matching of $P$. This is a graph over $P$, where $(p,q)$ is 
an edge if and only if the number of points of $P$ whose distance from both $p$ and $q$ is smaller 
than $\dist(p,q)$ is at most 16. As shown in~\cite{CTL}, the number of edges in $\RNG17(P)$ is $O(n)$, 
so we could use the set of distances corresponding to these edges as our candidate set $D$. However, 
we do not know how to compute $\RNG17(P)$ efficiently. The best known algorithm for computing the $k$RNG, 
for any fixed integer $k$, of a set of $n$ points, is due to Su and Chang~\cite{SuC91}, and it runs in 
$O(n^{5/3}\log n)$ time (with the constant of proportionality depending on $k$).
This bound can be improved, using more recent range searching techniques,\footnote{%
  The best improvement we have found runs in $O^*(n^{7/5})$ time.}
but it seems unlikely that such an improvement will be near the bound we are after.

\begin{figure}[htb]
\begin{center}
\includegraphics{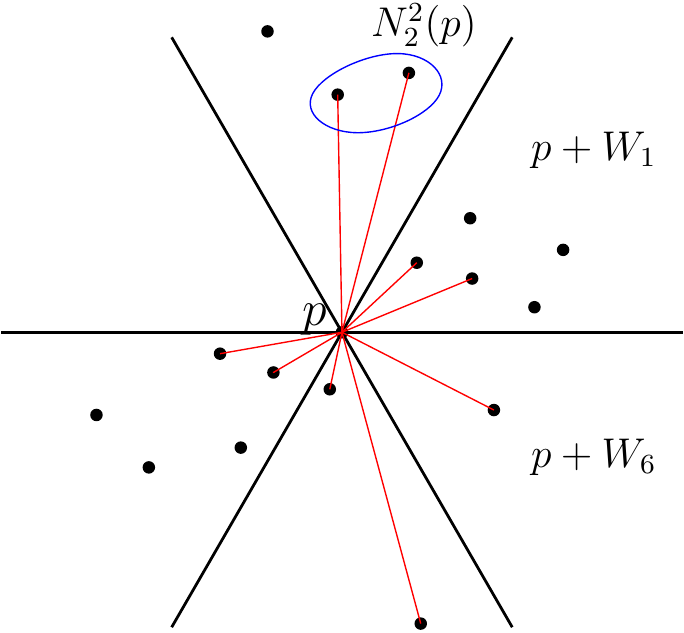}
\caption{The wedges $p+W_1, \ldots, p+W_6$ and the edges that are added to 2GNG$(P)$ `due' to $p$.} 
\label{fig:17gng}
\end{center}
\end{figure}

Instead, we consider another graph, known as the \emph{17-geographic neighborhood graph} of $P$, 
denoted as $\GNG17(P)$, which has also been studied in \cite{CTL} and is defined as follows. 
Partition the plane into six closed wedges $W_1,\ldots,W_6$, each with opening angle $60^\circ$, 
by the three lines through the origin with slopes $0$, $\pm \sqrt{3}$. The graph $\GNG17(P)$ 
consists of all the edges $(p,q)\in P\times P$, with $p\ne q$, for which there exists $1\le i\le 6$
such that $q\in p+W_i$, where $p+W_i$ is $W_i$ translated by $p$, so that there are at most $16$ 
points $q'\in p+W_i$ with $\dist(p,q') < \dist(p,q)$. (When the slope of the line through $p$ and 
$q$ is in $\{0,\pm\sqrt{3}\}$ there are two such wedges $W_i$.)
See Figure~\ref{fig:17gng} for an illustration. 
As shown in~\cite{CTL}, $\GNG17(P)$ is a supergraph of $\RNG17(P)$, and the number of its edges is 
still $O(n)$. We show below that $\GNG17(P)$ can be computed in $O(n\log^3n)$ time. Once we have computed 
$\GNG17(P)$, we perform a binary search on the set of distances corresponding to its edge set, 
as described above, to obtain $M^*$, in total time $O(n^{\omega/2}\log n)$. 

Before continuing, we remark that there is actually no need to compute $\GNG17(P)$. All we need is
the set of lengths of its edges, over which we will run binary search to find $M^*$ using the algorithm
of~\cite{BCM}. Nevertheless, for the sake of completeness, and since this is a result of independent 
interest, we present an algorithm that actually constructs $\GNG17(P)$ efficiently. The algorithm is 
relatively simple when the points of $P$ are in general position, but is somewhat more involved when 
this is not the case.

\paragraph{Computing $\GNG17(P)$ in general position.}
Begin by assuming that $P$ is in general position. Actually, the only condition that we require 
is that no three points form an isosceles triangle. Computing $\GNG17(P)$ in this case is 
relatively simple, since informally, the ranges that arise in computing $\GNG17(P)$ are 
wedges with fixed orientations of their bounding rays. In contrast, the ranges that arise in 
computing $\RNG17(P)$ are lenses formed by the intersection of two congruent disks, which are
more expensive to process.

\begin{figure}[htb]
\begin{center}
\includegraphics{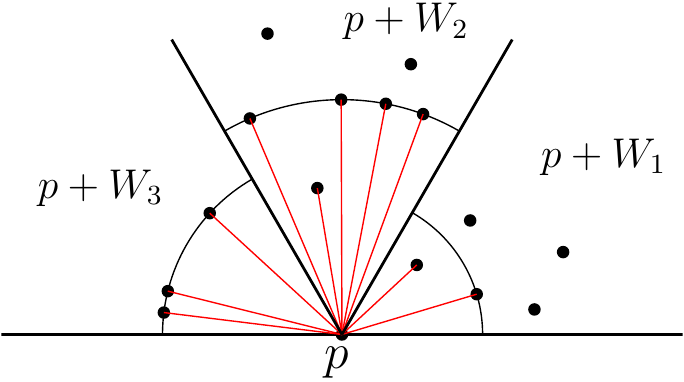}
\caption{When $P$ is not in general position $|N_i^2(p)|$ could be larger than 2. 
Here, $|N_1^2(p)|=2$, $|N_2^2(p)|=5$, and $|N_3^2(p)|=3$.} 
\label{fig:degeneracies}
\end{center}
\end{figure}

For each $p \in P$, we compute the set $N_i^{17}(p)$, for $i = 1,\ldots,6$, which
consists of the points that have at most 16 other points in $P\cap (p+W_i)$ that are closer to $p$.
When $P$ is in general position, these are the 17 closest points to $p$ in $P\cap (p+W_i)$, 
see Figure~\ref{fig:17gng}. However, when this is not the case, $|N_i^{17}(p)|$ could be much larger,
see Figure~\ref{fig:degeneracies}. (If $|P\cap (p+W_i)| \le 17$, then $N_i^{17}(p)=P\cap (p+W_i)$.)
We will later discuss the extension of the algorithm and analysis to degenerate situations.  

We construct six similar data structures, where the $i$'th structure is used 
to compute the collection of sets $\{N_i^{17}(p) \mid p \in P\}$. The $i$'th structure is an 
augmented two-level `orthogonal' range searching structure, where each level stores the points of 
(suitable canonical subsets of) $P$ in their order in the direction perpendicular to a bounding 
ray of $W_i$. For each node $v$ in the second level of the structure, we compute the 17th-order
Voronoi diagram $\Vor17(Q_v)$ of $Q_v$, the canonical set of $v$, and store it at $v$ as a point-location
structure. (Recall that $\Vor17(Q_v)$ partitions the plane into maximal regions, each with a fixed
\emph{set} of 17 closest sites.) The total size of the data structure is thus $O(n\log^2 n)$, and it 
can be computed in total time $O(n\log^3 n)$, where the Voronoi diagrams are computed by the algorithm of 
Chan and Tsakalidis~\cite{CT}, which takes $O(|Q_v|\log |Q_v|)$ deterministic time per diagram.
(See also the earlier algorithm of Agarwal et al.~\cite{ABMS}, which is randomized and slightly less efficient.)  
In fact, since 17 is a constant, we can compute the diagram in a simpler incremental manner, starting 
with the standard diagram and adding one index at a time. This also takes
$O(|Q_v|\log |Q_v|)$ time per diagram.

Now, for each $p \in P$, we perform a query in the $i$'th structure to obtain the set $N_i^{17}(p)$.
We initialize this set to be empty. We then identify the $O(\log^2 n)$ second-level nodes, such 
that $P\cap (p+W_i)$ is the disjoint union of their canonical sets. Next, we visit these nodes, 
one at a time, and at each node $v$ we perform a query with $p$ in $\Vor17(Q_v)$, and update 
$N_i^{17}(p)$ accordingly, in constant time. A query thus takes $O(\log^3 n)$ time for each $p\in P$, 
for a total of $O(n\log^3 n)$ time. (The latter bound follows since, in general position, 
each set $N_i^{17}(p)$ is of constant size (at most $17$)). The overall output of this procedure 
is the desired graph $\GNG17(P)$.

\paragraph{Handling degeneracies.}
When $P$ is not in general position, some modifications are required. 
As already observed, in this case $|N_i^{17}(p)|$ might be much larger than 17 (it could be even 
as large as $n-1$), although the total number of edges in $\GNG17(P)$ remains $O(n)$~\cite{CTL}.
The problematic situation is when, for some $p\in P$ and some $1\le i\le 6$,
and for some canonical set $Q$ of $P\cap (p+W_i)$, 
there exists a large subset $Q'$ of $Q$, all of whose points lie on a circle centered at $p$, with 
at most 16 points of $Q$ inside the circle. In that case all points of $Q'$ form with $p$ potential 
edges of $\GNG17(P)$, but to retain the aforementioned running time, we cannot afford to spend more 
than $O(\log n)$ time while processing $Q$ (when querying with $p$), unless they are indeed edges of 
$\GNG17(P)$, in which case we may spend additional $O(|Q'|)$ time. However, at this point we do not know 
whether they are edges of $\GNG17(P)$, because other canonical sets might generate with $p$ shorter
edges that exclude the potentially numerous longest edges involving $Q$ from the graph.\footnote{%
  This issue disappears if we only maintain the lengths of the edges of $\GNG17(P)$, as
  all these longest edges have the same length.}

To address this issue, we modify the query with $p$ in the $i$'th structure as follows. 
As before, we first identify the $O(\log^2 n)$ second-level nodes of the structure, 
such that $P\cap (p+W_i)$ is the disjoint union of their canonical sets. The next stage 
consists of two rounds, where in each round we process each of these $O(\log^2 n)$ nodes. 
The sole purpose of the first round is to find the distances $d_1 < d_2 < \cdots < d_k$, 
where $k \le 17$, between $p$ and the $17$ closest points to $p$ among the points in 
$P \cap (p+W_i)$. (We write $k \le 17$, since the same distance may be attained by several points.) More precisely, if we were to sort the points in $P \cap (p+W_i)$ 
by their distance from $p$, resolving ties arbitrarily, then $d_1,\ldots,d_k$ would be
the distinct distances corresponding to the first 17 points in the sequence.
We now describe how to implement the first round in total $O(\log^3 n)$ time, 
for each fixed $p$ and $i$, that is, in overall $O(n\log^3 n)$ time.

We use the following notation. For a (two-dimensional open) cell $C$ of $\Vor17(Q_v)$, 
for some second-level node $v$, denote by $P(C)$ the set of the (uniquely defined) 17 
closest points of $Q_v$ to any point in $C$. 
Moreover, for any point $p$ in the closure of $C$, 
put $d_p(C) = \max \{\dist(p,q) \mid q \in P(C)\}$.

For fixed $p$ and $i$, we maintain a sorted sequence $A = A_{p,i}$ of the 17 closest points to $p$
among the points of $P\cap(p+W_i)$ that were encountered so far; initially $A$ is empty. 
(Ties are broken arbitrarily, and we limit the length of $A$ to 17 even when there are 
additional points, not in $A$, whose distance from $p$ is equal to that between $p$ and 
the last point in $A$.) At each second-level node $v$ of the structure, we locate $p$ in 
$\Vor17(Q_v)$, and proceed as follows. Let $C$ be a cell of the diagram such that $p$ lies 
in the closure of $C$ (the cell is not unique when $p$ lies on an edge or is a vertex of the 
diagram). Obtaining $C$ from the point location query is straightforward in all cases, and takes constant time. We update $A$ (or initialize it, if $v$ is the first visited node) by examining 
the points in $P(C)$. Specifically, we sort the set $P(C)$ by distance to $p$, and
either copy the resulting sequence into $A$, when $v$ is the first processed node, 
or otherwise merge it with $A$, retaining only the first 17 elements. 

Put $\Delta_{p,i} = \{\dist(p,q) \mid q\in A\}$, and note that $|\Delta_{p,i}| \le 17$, 
where a strict inequality is possible. We have:
\begin{lemma} \label{lem:17}
After processing all the $O(\log^2n)$ canonical sets that form $P\cap(p+W_i)$, $\Delta_{p,i}$ 
is the set of distances between $p$ and the $17$ closest points to $p$ in $P \cap (p+W_i)$. 
\end{lemma}
\noindent{\bf Proof.}
We argue, by induction, that after processing some of the canonical sets, $\Delta_{p,i}$
is the set of distances between $p$ and the 17 closest points to $p$ in the union 
of these sets. The claim holds trivially initially, so assume that it holds
just before processing some canonical set $Q$.

Let $C$ be the cell of $\Vor17(Q)$ that is obtained from the point location query, so
$p$ lies in the closure of $C$. The claim is easy to argue when $p$ lies in the interior 
of $C$, so the interesting case is when $p\in\bd C$. We observe that in this case, if $C'$ 
is any other cell that contains $p$ (necessarily on its boundary) then $d_p(C') = d_p(C)$ and 
\[
P(C') \setminus \{q \in P(C') \mid \dist(p,q) < d_p(C')\} = 
P(C) \setminus \{q \in P(C) \mid \dist(p,q) < d_p(C)\} .
\]
This follows directly from the definition of the diagram, since otherwise, either $P(C)$ 
or $P(C')$ (or both) would not be the correct answer to the query with $p$.
In other words, in this case it does not matter which of the cells that are adjacent 
to $p$ we use to update $A$. This property implies that the invariant is preserved 
after processing $Q$, and it therefore completes the induction step in this case too.
$\Box$

In the second round,\footnote{%
  The second round is not needed if we are interested only in the lengths of the edges of $\GNG17(P)$.}
we iterate again over all points $p$ and indices $i$. For a fixed
pair $p$ and $i$, we retrieve the $O(\log^2n)$ canonical sets that form $P\cap (p+W_i)$.
For each of these sets $Q$, we locate $p$ in $\Vor17(Q)$ and compute $d_p(C)$ in constant time, 
for some (single) cell $C$ of the diagram whose closure contains $p$ (recall that this value is
independent of $C$). We then compute the set $\Delta_{p,i}(C)$ of distances between $p$ and the 
points of $P(C)$ that belong to $\Delta_{p,i}$, i.e., 
$\Delta_{p,i}(C) = \{d \in \Delta_{p,i} \mid d = \dist(p.q) \text{ for some } q \in P(C)\}$.
Two cases can arise:

\medskip
\noindent
(i) $d_p(C) \notin \Delta_{p,i}(C)$. In this case the subset $P_p(C)$ of $P(C)$ of 
the points whose distances to $p$ are in $\Delta_{p,i}(C)$ is independent of the cell $C$ (as 
long as $C$ contains $p$ in its closure). We then simply find $P_p(C)$ and add 
to $\GNG17(P)$ all the (at most 16) edges $(p,q)$ for $q\in P_p(C)$.

\medskip
\noindent
(ii) $d_p(C) \in \Delta_{p,i}(C)$. In this case we iterate over \emph{all} the cells
$C$ incident to $p$, and add all the edges $(p,q)$, for $q\in\bigcup_{C} P(C)$.
This does not affect the asymptotic bound on the running time: if there are $k$ 
such cells, we process $O(k)$ points but add $\Theta(k)$ edges to $\GNG17(P)$, 
as is easily checked. (If $p$ is adjacent to $k$ cells then there are $k$ points in $Q$ at distance $d_p(C)$ from $p$,
and all of them form with $p$ edges of $\GNG17(P)$.)

By applying this procedure to each $p$ and $i$ and to each corresponding canonical set $Q$,
we obtain the desired graph $\GNG17(P)$.


\paragraph{Wrapping it up.}
We have shown that one can compute $\GNG17(P)$ in $O(n\log^3 n)$ time.
Having computed $\GNG17(P)$, we continue as before, running a binary search over its $O(n)$ edge 
lengths, using the algorithm of Bonnet et al.~\cite{BCM} to guide the search.
In summary, we obtain:

\begin{theorem}
A bottleneck matching of a set of $n=2\ell$ points in the plane can be computed in $O(n^{\omega/2}\log n)$
deterministic time.
\end{theorem}

Abu-Affash et al.~\cite{AbuACKT} presented an algorithm that, given a perfect matching $M$ of $P$, 
returns in $O(n\log n)$ time a \emph{non-crossing} perfect matching $N$ of $P$, such that 
$\lambda(N) \le 2\sqrt{10}\lambda(M)$. We thus obtain:
 
\begin{corollary}
A non-crossing perfect matching $N$ of a set of $n=2\ell$ points in the plane with 
$\lambda(N) \le 2\sqrt{10}\lambda^*$ can be computed in $O(n^{\omega/2}\log n)$ deterministic 
time, where $\lambda^*$ is the length of a longest edge in a bottleneck matching of $P$.   
\end{corollary}


\begin{thebibliography}{}

\bibitem{AbuACKT}
A. K. Abu{-}Affash, P. Carmi, M. J. Katz and Y. Trabelsi,
Bottleneck non-crossing matching in the plane,
{\it Comput. Geom.} 47 (2014), 447--457.

\bibitem{AASS}
P. K. Agarwal, B. Aronov, M. Sharir and S. Suri,
Selecting distances in the plane,
{\it Algorithmica} 9 (1993), 495--514.

\bibitem{ABMS}
P. K. Agarwal, M. de Berg, J. Matou\v{s}ek and O. Schwarzkopf,
Constructing levels in arrangements and higher-order Voronoi diagrams,
{\it SIAM J. Comput.} 27 (1998), 654--667.

\bibitem{AVW}
J. Alman and V. Vassilevska Williams,
A refined laser method and faster matrix multiplication,
{\it Proc. ACM-SIAM Sympos. on Discrete Algorithms (SODA)}, 2021, 522--539.

\bibitem{BCM}
\'E. Bonnet, S. Cabello and W. Mulzer,
Maximum matchings in geometric intersection graphs,
{\it Proc. 37th Internat. Sympos. on Theoretical Aspects of Computer Science (STACS)}, 2020, 31:1--31:17. 
See also, arXiv:1910.02123.

\bibitem{CT}
T. M. Chan and K. Tsakalidis,
Optimal deterministic algorithms for 2-d and 3-d shallow cuttings,
{\it Discrete Comput. Geom.} 56 (2016), 866--881.

\bibitem{CTL}
M. S. Chang, C. Y. Tang and R. C. T. Lee,
Solving the Euclidean bottleneck matching problem by $k$-relative neighborhood graphs,
{\it Algorithmica} 8 (1992), 177--194.

\bibitem{CCJ}
B. N. Clark, C. J. Colbourn and D. S. Johnson,
Unit disk graphs,
{\it Discrete Math.} 86 (1990), 165--177.

\bibitem{GT}
H. N. Gabow and R. E. Tarjan,
Algorithms for two bottleneck optimization problems,
{\it J. Algorithms} 9 (1988), 411--417.

\bibitem{LeG}
F. Le Gall,
Powers of tensors and fast matrix multiplication,
{\it Proc. 39th Internat. Sympos. on Symbolic and Algebraic Computation (ISSAC)}, 2014, 296--303.

\bibitem{SuC91}
T.-H. Su and R.-C. Chang,
Computing the \emph{k}-relative neighborhood graphs in Euclidean plane,
{\it Pattern Recognition} 24 (1991), 231--239.

\end{thebibliography}
\end{document}